\documentclass[3p,times,twocolumn]{elsarticle}

\usepackage{ecrc}

\volume{00}

\firstpage{1}

\journalname{Nuclear Physics B Proceedings Supplement}

\runauth{K. Griessinger}

\jid{nuphbp}

\jnltitlelogo{Nuclear Physics B Proceedings Supplement}

\usepackage{amssymb}
\usepackage{amsmath}

\usepackage[figuresright]{rotating}
\usepackage{siunitx}
\usepackage[utf8]{inputenc}

\input{babarsym_mod.tex}

\newcommand\amunew{\num{17.9}}
\newcommand\amunewstaterr{\num{0.1}}
\newcommand\amunewsysterr{\num{0.6}}

\newcommand\amuhinew{\num{17.4}}
\newcommand\amuhinewstaterr{\num{0.1}}
\newcommand\amuhinewsysterr{\num{0.6}}

\newcommand\amuwidenew{\num{21.8}}
\newcommand\amuwidenewstaterr{\num{0.1}}
\newcommand\amuwidenewsysterr{\num{0.7}}

\begin{document}

\begin{frontmatter}



\dochead{}

\title{New ISR Cross Section Results on $e^+e^- \to \pi^+\pi^-\pi^0\pi^0$ and $e^+e^- \to \pi^+\pi^-\eta$ from \babar}


\author{Konrad Griessinger}

\address{Institute for Nuclear Physics, Mainz University}

\begin{abstract}
  Two new hadronic cross sections measured by the \babar experiment are presented: $e^+e^- \to \pi^+\pi^-\pi^0\pi^0$ and $e^+e^- \to \pi^+\pi^-\eta$. For both channels, the contribution the anomalous magnetic moment of the muon is calculated.
\end{abstract}


\end{frontmatter}



\section{Introduction}\label{sec:intro}

The muon gyromagnetic factor $g_\mu$ can be determined directly from spin precession measurements and theoretically in the Standard Model. 
In the latter, the leading order hadronic part needs to be calculated from the experimentally determined hadronic cross section $\sigma_\mathrm{had}(s)$ using a relation derived from the optical theorem and a dispersion integral:


\begin{equation*}
a_\mu^\mathrm{had} = \frac{1}{4\pi^3} \int_{m_\pi^2}^\infty K_\mu(s) \cdot \frac{\sqrt{1 - \frac{4m_e^2}{s}}}{1 + \frac{2m_e^2}{s}} \cdot \sigma^{(0)}_\mathrm{had}(s) \mathrm{d}s
\text{ .}
\end{equation*}


$K(s)$ denotes the analytically known Kernel function~\cite{Achasov:2002bh}, while $\sigma^{(0)}$ is the Born cross section measured in $e^+e^-$ collisions.
The hadronic higher order contributions from vacuum polarization and light-by-light scattering are considered elsewhere (e.g.~\cite{Jegerlehner:2009ry,Kurz:2014wya,Colangelo:2014qya} and references therein).


The direct and theoretical results for $a_\mu$ deviate by more than $\num{3} \sigma$. 
Since this discrepancy merely provides a hint at yet no evidence for a deficiency in the Standard Model, more precise data from both sides is necessary. Below, two recent \babar analyses of hadronic channels are presented, which significantly improve their respective contributions to $a_\mu$.

\section{Experimental Setup}\label{sec:expset}

The \BF \babar at SLAC National Accelerator Laboratory in Stanford, USA achieved an integrated luminosity of $\sim \SI{518}{fb^{-1}}$~\cite{Bevan:2014iga}. While running (mostly) at the $\Upsilon(4S)$ resonance, this produced a large data set for the study of hadronic cross sections via the process of Initial State Radiation (ISR). This process 
takes place when one of the particles in the initial state ($e^+e^-$) radiates a photon. Since the photon carries a certain amount of CMS-energy $E_\gamma^\ast$, the squared center-of-mass energy $s$ of the collision ist lowered to $s^\prime = s - 2 \sqrt{s} E_\gamma^\ast$. Owed to the continuous spectrum of photon energies, this gives access to a broad range of invariant masses over which hadronic cross sections may be measured. 


\section{The channel $e^+e^- \to \pi^+\pi^-\pi^0\pi^0$}\label{sec:2pi2pi0}

To date, the cross section $e^+e^- \to \pi^+\pi^-\pi^0\pi^0$ was one of the contributions limiting the precision of $a_\mu^\mathrm{had}$ since especially at energies above $\SI{1.4}{GeV}$ no precise measurement existed. Therefore an analysis up to an energy of \SI{4.5}{GeV} is performed on the full \babar data set using the ISR method.

The most important event selection requirements used in this analysis are
\begin{itemize}
\item exactly 2 charged tracks,
\item $\geq 5$ photons,
\item $E^\mathrm{lab}_{\gamma} > \SI{0.05}{GeV}$,
\item $| M^\mathrm{reco}_{\pi^0} - M^\mathrm{PDG}_{\pi^0}| < \SI{0.03}{GeV}$,
\item kinematic fit: $\chi^2_{2\pi2\pi^0\gamma} < 30$.
\end{itemize}

\begin{figure}
  \centering
  \includegraphics[type=pdf,ext=.jpg,read=.jpg,trim = 0 0 0 0, clip, width=\linewidth]{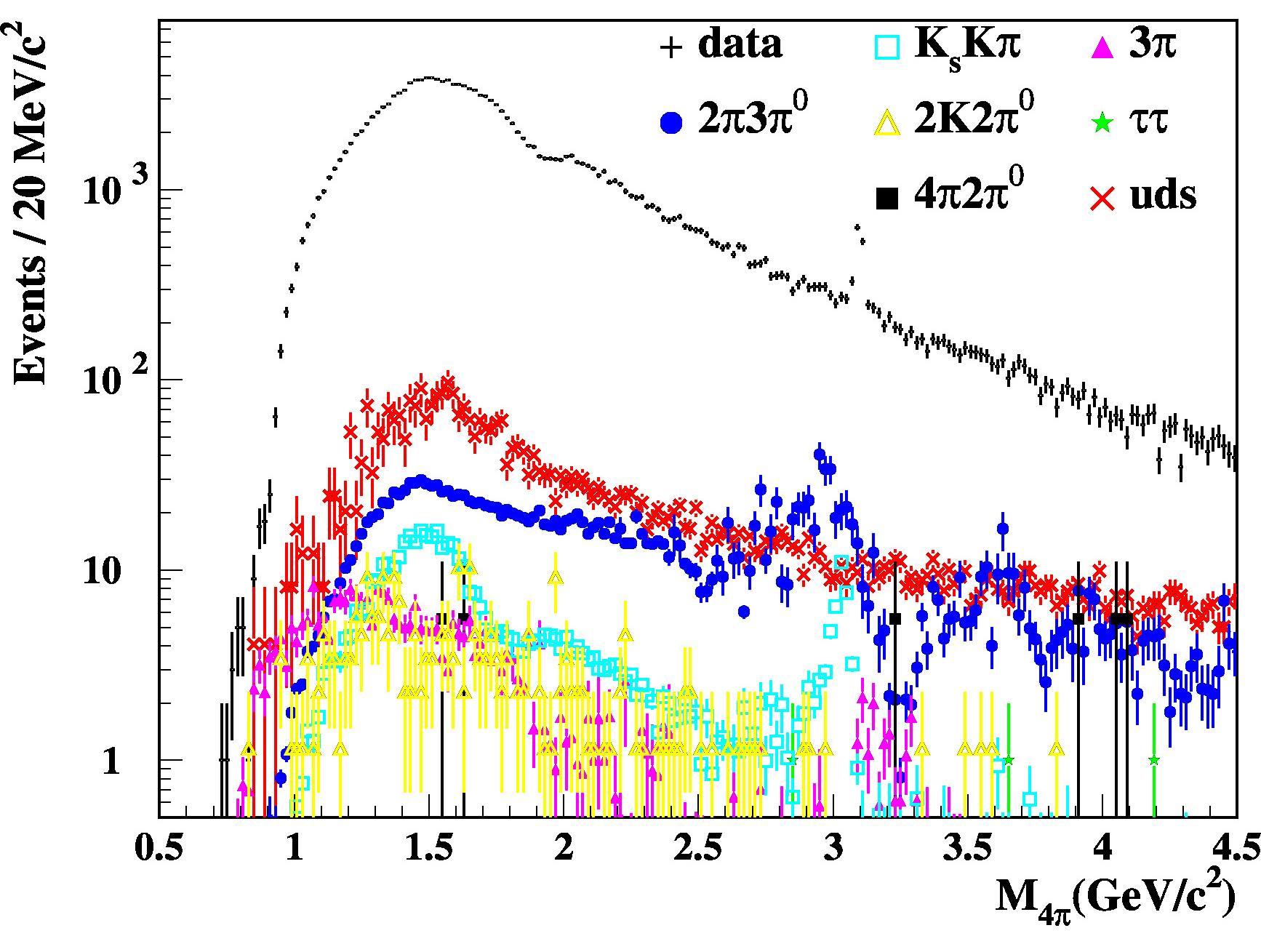}
  \caption{The backgrounds to $\pi^+\pi^-2\pi^0$ data: uds-continuum, $\pi^+\pi^-\pi^0$, $2(\pi^+\pi^-\pi^0)$, $K_\mathrm{s}K^\pm\pi^\mp$, $K^+K^-2\pi^0$, $\tau^+\tau^-$, and $\pi^+\pi^-3\pi^0$ as a function of $M_{4\pi}$.}
  \label{fig:allbkg}
\end{figure}
\begin{figure}
  \centering
  \includegraphics[trim = 0 0 0 0, clip, width=\linewidth]{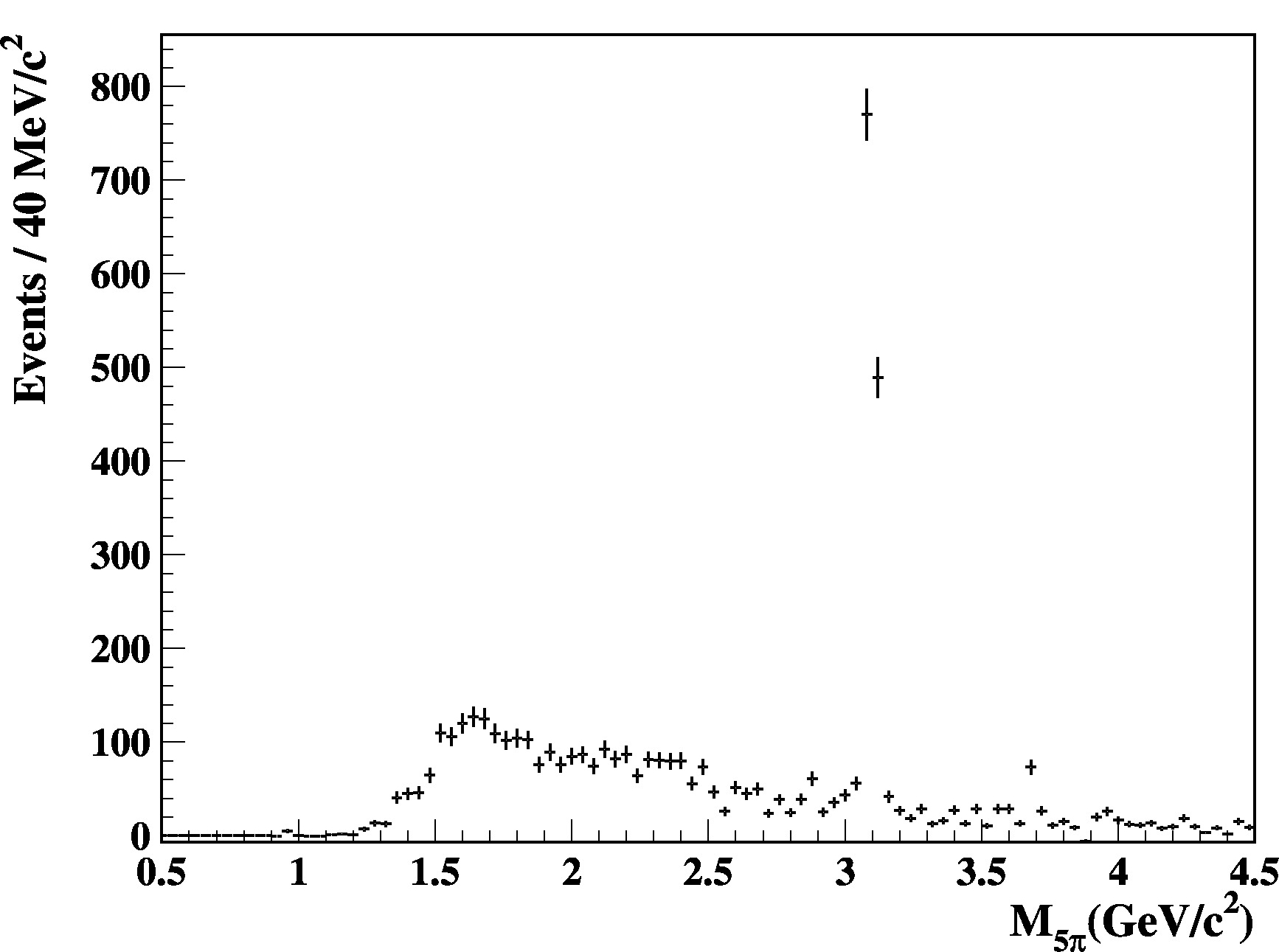}
  \caption{Measured $e^+e^- \to \pi^+\pi^-3\pi^0$ count rate.}
  \label{fig:5pirate}
\end{figure}

A small fraction of background events survives the selection requirements as shown in Fig.~\ref{fig:allbkg}. These events are subtracted using Monte Carlo simulation. This is not possible for the channel $e^+e^- \to \pi^+\pi^-3\pi^0$ since the full process has never been measured precisely, hence no Monte Carlo generator exists for this process. Using the Monte Carlo samples of two subprocesses ($e^+e^- \to \omega2\pi^0 \to \pi^+\pi^-3\pi^0$ and $e^+e^- \to \eta\pi^+\pi^- \to \pi^+\pi^-3\pi^0$) for the efficiency determination, the count rate of the channel $e^+e^- \to \pi^+\pi^-3\pi^0$ was measured from \babar data. The result is shown in Fig.~\ref{fig:5pirate} and enables us to adjust the existing Monte Carlo samples to reflect the measured five pion mass distribution. The adjusted Monte Carlo sample is then used in the subtraction of backgrounds from the channel $e^+e^- \to \pi^+\pi^-2\pi^0$.

In order to cross check the simulation-based background subtraction, background is also subtracted using a data-driven sideband method.
Comparing the two background subtraction methods yields a discrepancy of less than \SI{1}{\%} in the peak region around \SI{1.5}{GeV/c^2}. 
Above \SI{2.7}{GeV/c^2} the discrepancy increases to approximately \SI{6}{\%}. The discrepancies between the different background subtraction methods serve as a measure for the corresponding systematic uncertainty.

Further systematic uncertainties are determined, especially concerning the $\pi^0$ detection efficiency, which gives an additional \SI{2}{\%} uncertainty over the full energy range. All other effects yield considerably smaller uncertainties.



Combining all systematic uncertainties in quadrature results in \SI{3.1}{\%} for $M_{4\pi}$ from \num{1.2} to \SI{2.7}{GeV/c^2}, \SI{6.7}{\%} for $M_{4\pi}$ from \num{2.7} to \SI{3.2}{GeV/c^2}, and \SI{7.1}{\%} above \SI{3.2}{GeV/c^2}. In the region below \SI{1.2}{GeV/c^2} the relative systematic uncertainty is mass-dependent.


\begin{figure}
  \centering
  \includegraphics[trim = 0 0 0 0, clip, width=\linewidth]{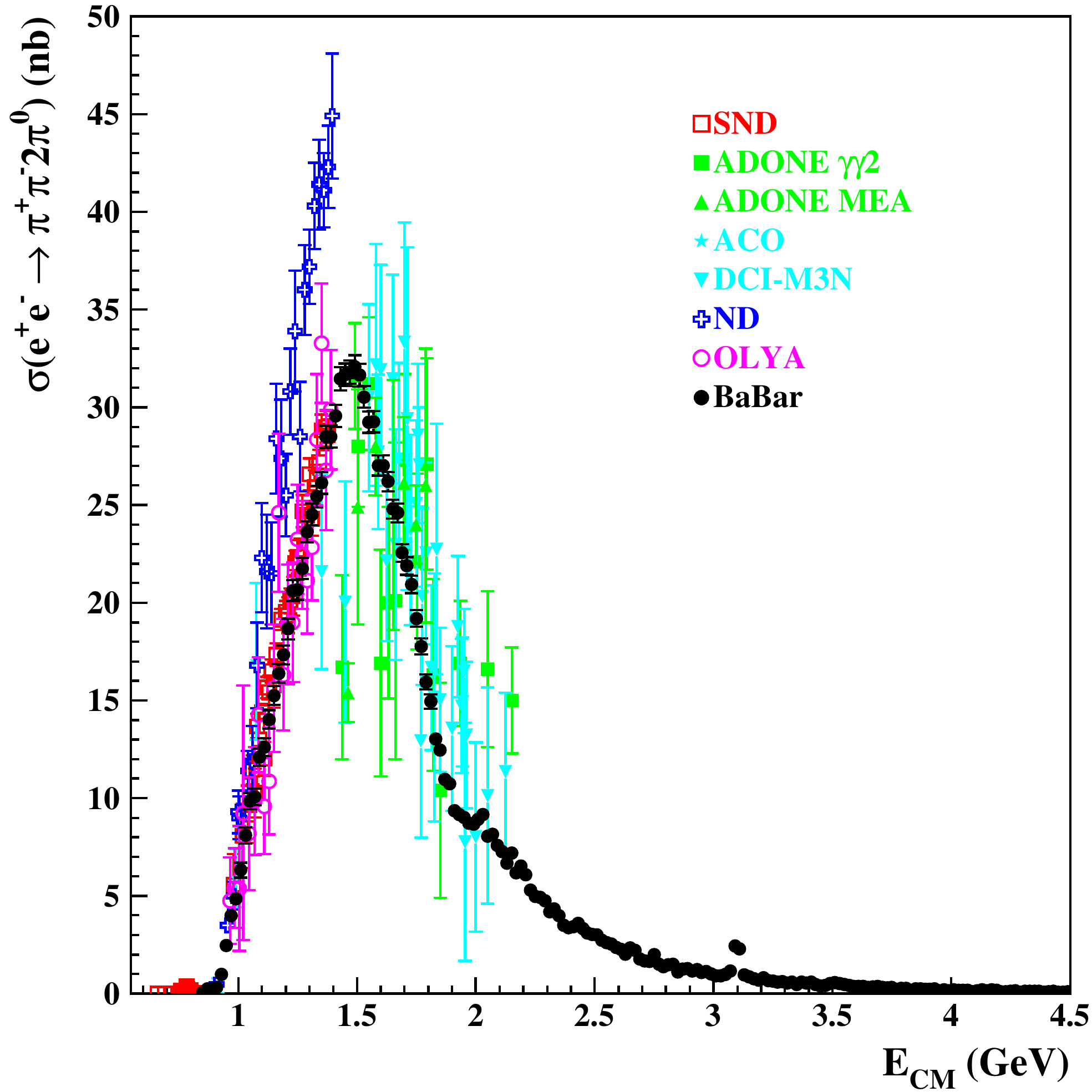}
  \caption{The cross section $e^+e^- \to \pi^+\pi^-2\pi^0$ compared to other measurements.}
  \label{fig:2pi2pi0csworld}
\end{figure}

\begin{figure}
  \centering
  \includegraphics[type=pdf,ext=.pdf,read=.pdf,trim = 0 0 0 50mm, clip, width=\linewidth]{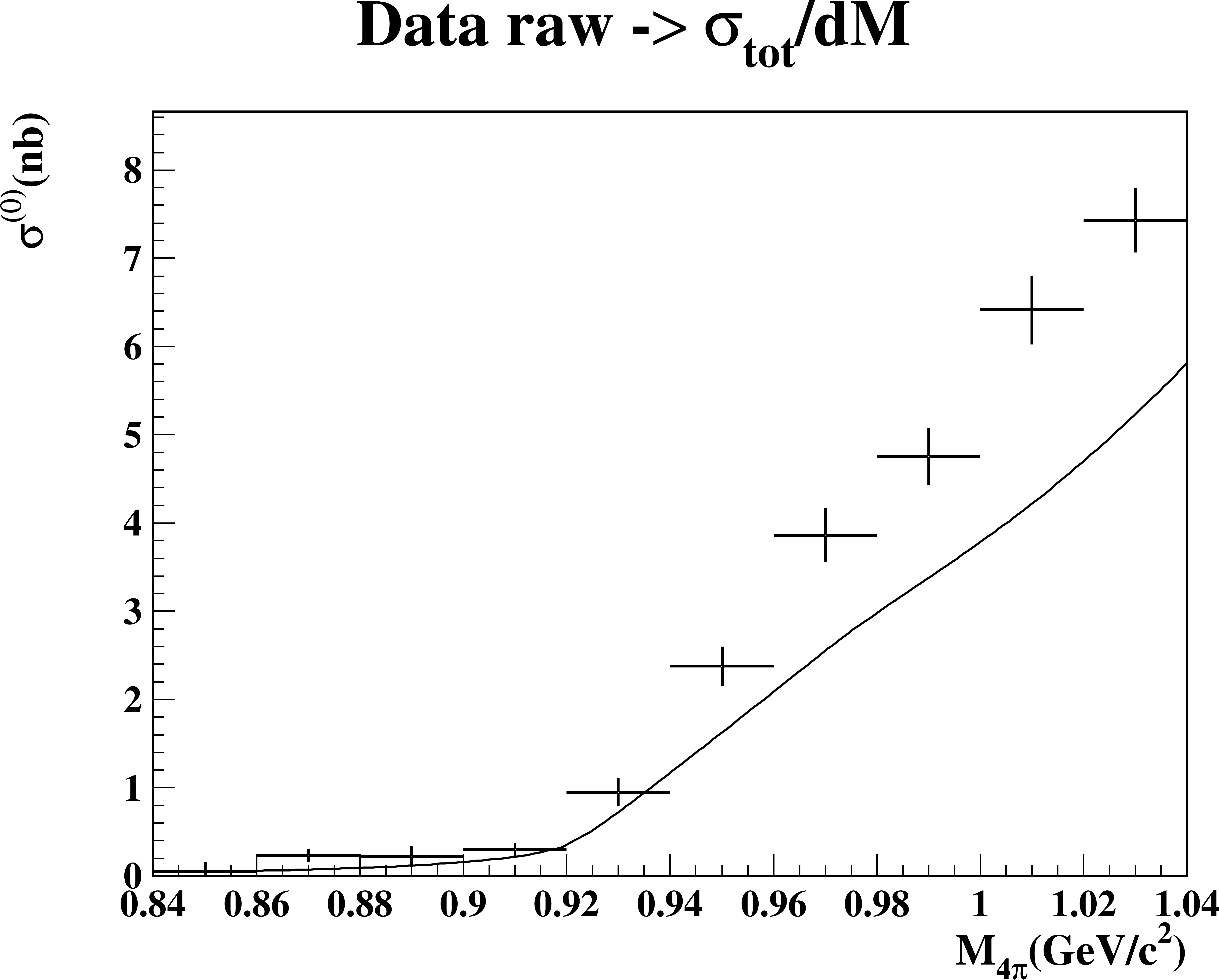}
  \caption{The cross section $e^+e^- \to \pi^+\pi^-2\pi^0$ in the low-mass region compared to theory.}
  \label{fig:2pi2pi0cslow}
\end{figure}

The cross section result is shown in Fig.~\ref{fig:2pi2pi0csworld} including statistical uncertainties in comparison to the existing data for this channel. A comparison to chiral perturbation theory~\cite{Ecker:2002cw} in the low-mass region is shown in Fig.~\ref{fig:2pi2pi0cslow}. 

From the new measurement an updated value of this channel's contribution to $a_\mu$ is extracted in the energy range $\num{0.85} < \sqrt{s} < \SI{1.8}{GeV}$:
\begin{equation*}
  a_\mu = (\amunew \pm \amunewstaterr_\mathrm{stat} \pm \amunewsysterr_\mathrm{syst}) \times 10^{-10} \text{ .}
\end{equation*}
In the wider range $\num{0.85} < \sqrt{s} < \SI{3.0}{GeV}$ the result is
\begin{equation*}
  a_\mu = (\amuwidenew \pm \amuwidenewstaterr_\mathrm{stat} \pm \amuwidenewsysterr_\mathrm{syst}) \times 10^{-10} \text{ .}
\end{equation*}
For comparison with existing data in the energy range $\num{1.02} < \sqrt{s} < \SI{1.8}{GeV}$, which resulted in $a_\mu=(\num{16.76} \pm \num{1.31} \pm \num{0.20}_\mathrm{rad}) \times 10^{-10}$~\cite{Davier:2003pw}, a new value in the same energy range is extracted, yielding $a_\mu = (\amuhinew \pm \amuhinewstaterr_\mathrm{stat} \pm \amuhinewsysterr_\mathrm{syst}) \times 10^{-10}$. This comparison shows that the new \babar measurement alone improves the uncertainty of the previously existing world data set by more than a factor \num{2}.

\section{The channel $e^+e^- \to \pi^+\pi^-\eta$}\label{sec:pipieta}

This channel was measured by \babar before~\cite{Aubert:2007ef} using the decay $\eta \to \pi^+\pi^-\pi^0$ on approximately half of the data set, leading to limited accuracy especially at energies above \SI{2}{GeV}. In the new measurement the decay $\eta \to \gamma\gamma$ is investigated on the full data set. This leads to much improved precision and an energy range reaching center-of-mass energies of \SI{3.5}{GeV}.



The main selection criteria for the analysis are
\begin{itemize}
\item at least 2 charged tracks,
\item $\geq 3$ photons,
\item $E^\mathrm{lab}_{\gamma} > \SI{0.1}{GeV}$,
\item $\num{0.44} < M^\mathrm{reco}_{\eta} < \SI{0.64}{GeV/c^2}$.
\end{itemize}

Non-peaking background is subtracted by approximating of two photon invariant mass while peaking background is subtracted using Monte Carlo samples normalized to data. 
The peaking backgrounds are $e^+e^- \to \pi^+\pi^-\pi^0\eta$, $e^+e^- \to \pi^+\pi^-\pi^0\eta\gamma$, and $e^+e^- \to K^+K^-\eta\gamma$.
The systematic uncertainties of the analysis are dominated by the background subtraction and amount to \SI{10}{\%} for $m_{\pi^+\pi^-\eta} < \SI{1.35}{GeV/c^2}$, \SI{4.5}{\%} for $\SI{1.35}{GeV/c^2} < m_{\pi^+\pi^-\eta} < \SI{1.80}{GeV/c^2}$, \SI{6.5}{\%} for $\SI{1.80}{GeV/c^2} < m_{\pi^+\pi^-\eta} < \SI{2.50}{GeV/c^2}$, \SI{11}{\%} for $\SI{2.50}{GeV/c^2} < m_{\pi^+\pi^-\eta} < \SI{3.10}{GeV/c^2}$, and \SI{12}{\%} for $m_{\pi^+\pi^-\eta} > \SI{3.10}{GeV/c^2}$.

\begin{figure}
  \centering
  \includegraphics[trim = 0 0 0 0, clip, width=\linewidth]{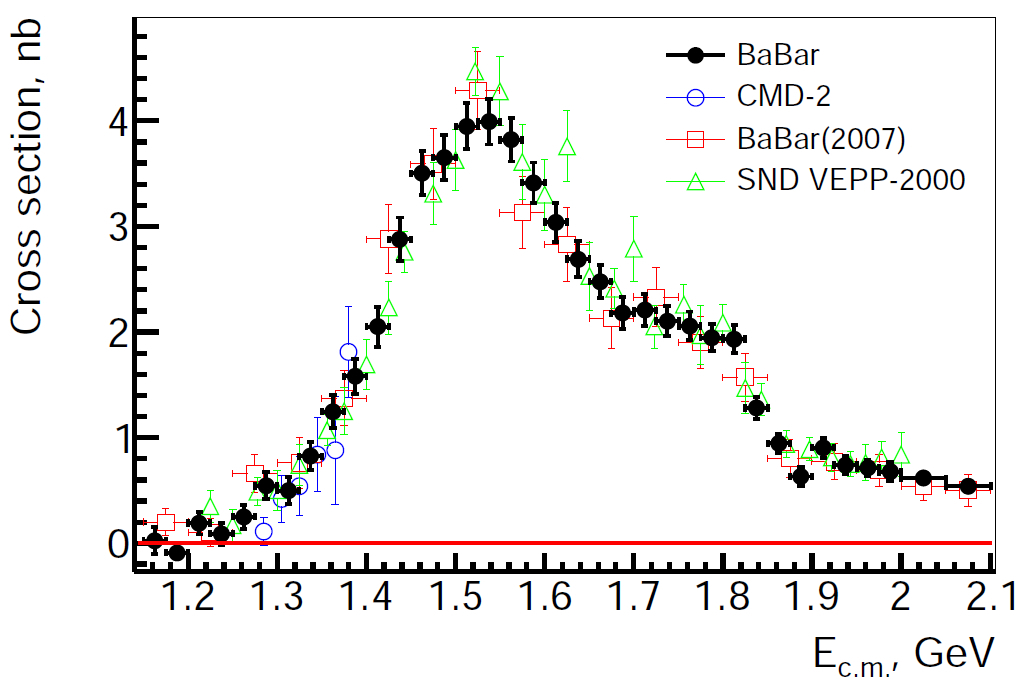}
  \caption{The cross section $e^+e^- \to \eta\pi^+\pi^-$ compared to other measurements.}
  \label{fig:pipietacsfull}
\end{figure}

\begin{figure}
  \centering
  \includegraphics[trim = 0 0 0 0, clip, width=\linewidth]{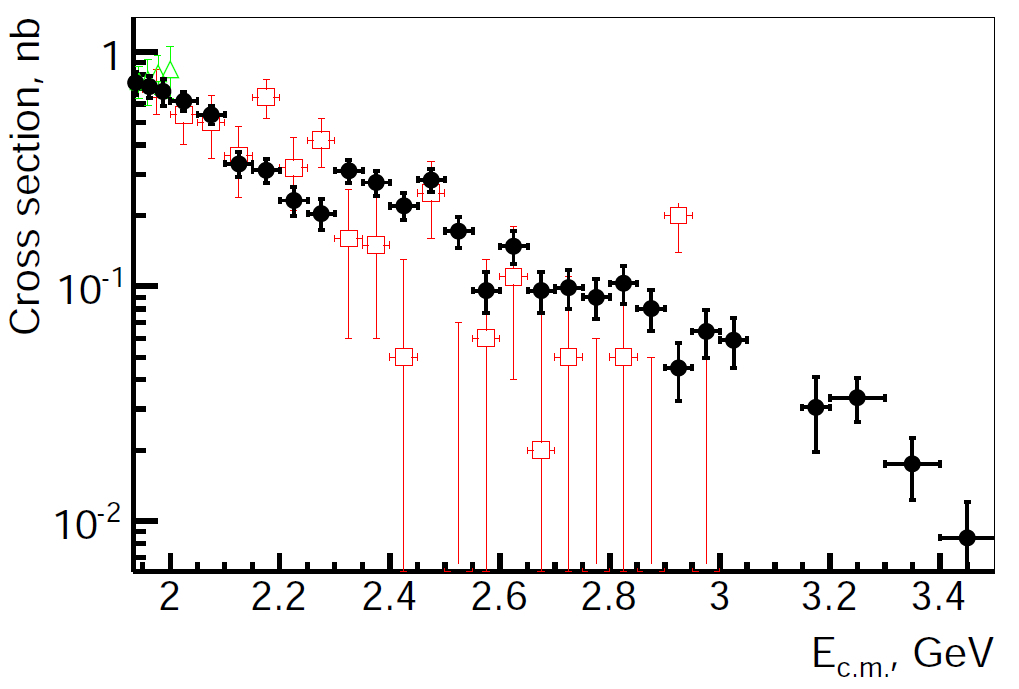}
  \caption{The cross section $e^+e^- \to \eta\pi^+\pi^-$.}
  \label{fig:pipietacslg}
\end{figure}

\begin{figure}
  \centering
  \includegraphics[trim = 0 0 0 0, clip, width=\linewidth]{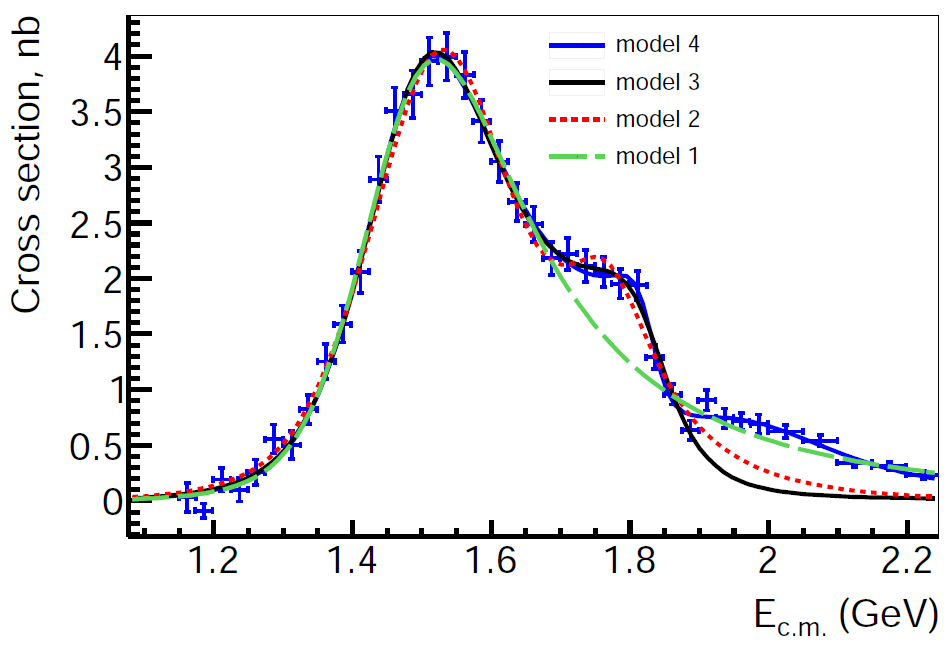}
  \caption{The cross section $e^+e^- \to \eta\pi^+\pi^-$.}
  \label{fig:pipietacsfit}
\end{figure}

After background subtraction and efficiency correction, the cross section shown in Fig.~\ref{fig:pipietacsfull} is extracted. It significantly improves the measurement precision compared to previously available data, especially in the high energy region shown in Fig.~\ref{fig:pipietacslg}. The cross section can be compared to several VMD models, which are fitted to the data. Four models are studied, where model 1 contains the resonances $\rho(770)$ and $\rho(1450)$ with relative phase \ang{180}. 
Model 2 contains the resonances $\rho(770)$ with phase \ang{0} and $\rho(1450)$ as well as $\rho(1700)$ with phase \ang{180}.
Model 3 contains the resonances $\rho(770)$ as well as $\rho(1700)$ with phase \ang{0} and $\rho(1450)$ with phase \ang{180}.
Model 4 contains the resonances $\rho(770)$, $\rho(1700)$, $\rho(2150)$ with phase \ang{0} and $\rho(1450)$ with phase \ang{180}.
Model 1 is fitted to data for energies $E_\mathrm{CM} < \SI{1.7}{GeV}$, while models 2 and 3 are fitted up to $E_\mathrm{CM} < \SI{1.9}{GeV}$ and for model 4 the energy range up to $E_\mathrm{CM} < \SI{2.2}{GeV}$ is used.
It is observed in Fig.~\ref{fig:pipietacsfit} that all fitted models describe data reasonably for energies below \SI{1.7}{GeV}. Model 1 deviates from data for higher energies and Models 2 and 3 fail above \SI{1.9}{GeV}, as expected from the fit ranges.

Using the newly measured cross section, the corresponding contribution to $a_\mu$ is extracted in the energy range below \SI{1.8}{GeV}:
\begin{equation*}
a_\mu = (1.19 \pm 0.02_\mathrm{stat} \pm 0.06_\mathrm{syst}) \times 10^{-10} \text{ .}
\end{equation*}
This new value is more precise than earlier results and may help resolve the tension between previously existing predictions yielding $(0.88 \pm 0.10) \times 10^{-10}$~\cite{Hagiwara:2011af} and $(1.15 \pm 0.06_\mathrm{stat} \pm 0.08_\mathrm{syst}) \times 10^{-10}$~\cite{Davier:2010nc}.


\section{Summary}\label{sec:summary}

The cross section of the process $e^+e^- \to \pi^+\pi^-2\pi^0$ is measured with a systematic precision of \SI{3.1}{\%} in its peak region. This leads to a new value of its contribution to the prediction of the anomalous magnetic moment of the muon $a_\mu = (\amunew \pm \amunewstaterr_\mathrm{stat} \pm \amunewsysterr_\mathrm{syst}) \times 10^{-10}$ in the energy range from \SI{0.85}{GeV} to \SI{1.8}{GeV}.

Furthermore, the cross section of the process $e^+e^- \to \eta\pi^+\pi^-$ is analyzed using the decay $\eta \to \gamma\gamma$. A systematic precision of \SI{4.5}{\%} is reached in its peak region, leading to a contribution to $a_\mu = (1.19 \pm 0.02_\mathrm{stat} \pm 0.06_\mathrm{syst}) \times 10^{-10}$ in the energy range from threshold to \SI{1.8}{GeV}.

\section{Acknowledgments}\label{sec:acknowledgments}
We would like to thank the organizers for this wonderful conference. 
We are grateful for the excellent luminosity and machine conditions
provided by our \pep2 colleagues, 
and for the substantial dedicated effort from
the computing organizations that support \babar.
The collaborating institutions wish to thank 
SLAC for its support and kind hospitality. 
This work is supported by
DOE
and NSF (USA),
NSERC (Canada),
CEA and
CNRS-IN2P3
(France),
BMBF and DFG
(Germany),
INFN (Italy),
FOM (The Netherlands),
NFR (Norway),
MES (Russia),
MICIIN (Spain),
STFC (United Kingdom). 
Individuals have received support from the
DFG (Germany).





\bibliographystyle{elsarticle-num}
\bibliography{procbib}







\end{document}